\documentclass[epj]{svjour}
\usepackage{amsfonts}
\usepackage{graphics}
\usepackage{graphicx}
\usepackage{amsmath}
\usepackage{amssymb,bbold}
\RequirePackage[numbers,sort&compress]{natbib}
\RequirePackage[colorlinks,citecolor=blue,urlcolor=blue,linkcolor=blue]{hyperref}
\RequirePackage{flushend}
\usepackage{subfig}
\journalname{EPJ A}

\begin{document}
\title{Missing solution in a relativistic Killingbeck potential}
\author{Luis B. Castro\inst{1}\and
Angel E. Obispo \inst{2}
}                     
%
%
\institute{Departamento de F\'{\i}sica, Universidade Federal do Maranh\~{a}o, Campus Universit\'{a}rio do Bacanga, 65080-805, S\~{a}o Lu\'{i}s, MA, Brazil, \email{lrb.castro@ufma.br,luis.castro@pq.cnpq.br} \and 
Departamento de F\'{\i}sica -- IGCE, Universidade Estadual Paulista (UNESP), Campus de Rio Claro, 13506-900, Rio Claro, SP, Brazil, \email{aeobispo@rc.unesp.br}
}
\date{Received: date / Revised version: date}
%
\abstract{
Missing bound--state solutions for fermions in the background of a Killingbeck radial potential including an external magnetic and Aharonov--Bohm (AB) flux fields are examined. The correct quadratic form of the Dirac equation with vector and scalar couplings under the spin and pseudo--spin symmetries is showed and also we point out a misleading treatment in the literature regarding to bound--state solutions for this problem.
} 
\maketitle

\section{Introduction}
\label{sec:introduction}

The study of low-dimensional fermion systems has long been recognized as
important in understanding various phenomena in different areas of physics,
such as Particle physics, Condensed matter physics, among others. One of the
main reasons is than it provides a simplified version of the
three-dimensional world. However, these systems are more than a prototype of
the higher-dimensional systems and are being scenario of new and exotics
phenomena that have generated a great interest, both theoretical and applied
physics. 

On the theoretical side, the fermion number (charge) fractionalization in
relativistic quantum field theory is a remarkable phenomenon and that was
shown to occur in one dimensional systems \cite{PRD13:3398:1976} (as
polyacetylene \cite{PRL42:1698:1979,PRB22:2099:1980,NPB190:253:1981}) when fermions interact with background
fields with a topologically nontrivial soliton profile. In this context, isolated zero modes (isolated solutions) of the fermion--soliton system can have
fractional fermion numbers of $\pm 1/2$. Similarly, the existence of these
isolated solutions in bidimensional condensed matter systems are also
responsible to induce a fractional charge. For example, in \cite%
{PRL98:186809:2007} the authors shown that when lattice distortions with
vortex profile are incorporated in a graphene lattice, there are zero modes
excitations in the single-particle energy spectrum. From the existence of
isolated zero modes and from the sublattice
symmetry, they show that the fermion quantum charge is fractionalized, which
even persists if \cite{PRL98:186809:2007} is extended to a chiral gauge
theory for graphene \cite{PRL98:266402:2007,PRB89:165405:2014,PRB91:035404:2015}. 

In this paper we consider some interactions used in \cite{PRL98:186809:2007,PRL98:266402:2007,PRB89:165405:2014,
PRB91:035404:2015}. Particularly, we study the dynamics of fermions
in $2+1$ dimensions under a influence of mixture of scalar $S(r)$, vector $%
V(r)$ and minimal $\vec{A}(r)$ interactions. The classification of the
potentials are based on the behavior under a Lorentz transformation: $S(r)$
for the Lorentz scalar, and $V(r)$ and $\vec{A}(r)$ for the time and space
components of a two--vector potential, respectively. Each of these
interactions has important applications and the study of its effects on the
dynamics of fermions are of great interest in the scientific community \cite%
{STRANGE1998}. The time and space components of a two--vector potential is useful for
studying the dynamics of a spin--$1/2$ charged particle in an electric and
magnetic fields, respectively. On the other hand, the scalar potential can
be interpreted as a position--dependent mass.

The case in which the
couplings are composed by a vector $V(r)$ and a scalar $S(r)$ potentials,
with $S(r)=V(r)$ [or $S(r)=-V(r)$], are usually pointed out as necessary
condition for occurrence of exact spin [or pseudo--spin] symmetry. It is
known that the spin and pseudo--spin symmetries are SU(2) symmetries of a
Dirac Hamiltonian with vector and scalar potentials. The pseudo--spin
symmetry was introduced in nuclear physics many years ago \cite%
{PLB30:517:1969,NPA137:129:1969} to account for the degeneracies of orbital
in single-particle spectra. Also, it is known that the spin symmetry occurs
in the spectrum of a meson with one heavy quark \cite{PRL86:204:2001} and
anti-nucleon bound in a nucleus \cite{PR315:231:1999}, and the pseudo--spin
symmetry occurs in the spectrum of nuclei \cite{PRL78:436:1997}.

In a recent article in this journal \cite{EPJA52:201:2016}, Eshghi and collaborators investigated the Dirac equation in $2+1$ dimensions with a
Killingbeck radial potential including an external magnetic and
Aharonov--Bohm (AB) flux fields. They mapped the Dirac equation into
Sturm--Liouville problem of a Schr\"{o}dinger--like equation and obtained a
set of bound--state solutions by recurring to the properties of the
biconfluent Heun equation. Nevertheless, an isolated solution from the
Sturm--Liouville scheme was not taken into account. The purpose of this work
is to report on this missing bound--state solution and additionally we point
out a misleading treatment in Ref. \cite{EPJA52:201:2016}. Additionally, we
shed some light on a misconception recently propagated in the literature
with respect to Aharonov--Bohm (AB) potential.

\section{Dirac equation in $2+1$ dimensions}
\label{sec2}

The Dirac equation in $2+1$ dimensions in
polar coordinates is given by ($\hbar=c=1$)
\begin{equation}
\left\{ \beta \boldsymbol{\gamma }\cdot \mathbf{\pi }+\beta \left[ M+S\left(
r\right) \right] \right\} \psi \left( \mathbf{r}\right) =\left[ E-V\left(
r\right) \right] \psi \left( \mathbf{r}\right) ,  \label{dirac2}
\end{equation}%
where $\mathbf{\pi }=\left( \pi _{r},\pi _{\varphi }\right) =(-i\partial
_{r},-i\partial _{\varphi }/r-eA_{\varphi })$, $\mathbf{r}=(r,\varphi )$ and
$\psi $ is a two-component spinor. Here $E$ is the energy of the fermion, $S(r)$ is  a scalar potential, $V(r)$ is the time--like vector potential and $A_{\varphi}$ is the space--like vector potential. The $\boldsymbol{\gamma }$ matrices in
Eq. (\ref{dirac2}) are given in terms of the Pauli matrices as \cite%
{EPJC74:3187:2014}
\begin{align}
\beta \gamma ^{r}& =\sigma _{1}\cos \varphi +s\sigma _{2}\sin \varphi
=\left(
\begin{array}{cc}
0 & e^{-is\varphi } \\
e^{+is\varphi } & 0%
\end{array}%
\right),  \label{sgimar} \\
\beta \gamma ^{\varphi }& =-\sigma _{1}\sin \varphi +s\sigma _{2}\cos
\varphi =\left(
\begin{array}{cc}
0 & -ise^{-is\varphi } \\
ise^{+is\varphi } & 0%
\end{array}%
\right) ,  \label{sgimaphi} \\
\beta & =\sigma _{3}=\left(
\begin{array}{cc}
1 & 0 \\
0 & -1%
\end{array}%
\right) ,  \label{sgimaz}
\end{align}%
where $s$ is twice the spin value, with $s=+1$ for spin \textquotedblleft
up\textquotedblright\ and $s=-1$ for spin \textquotedblleft
down\textquotedblright . It is worthwhile to mention that the representations (\ref{sgimar}), (\ref{sgimaphi}) and (\ref{sgimaz}) are most suitable for $2+1$ dimensions. Equation (\ref{dirac2}) can be written more
explicitly as%
\begin{eqnarray}
e^{-is\varphi }\left[ \pi _{r}-is\pi _{\varphi }\right] \psi _{2} &=&\left[
E-M-\Sigma \left( r\right) \right] \psi _{1},  \label{dirac3a} \\
e^{+is\varphi }\left[ \pi _{r}+is\pi _{\varphi }\right] \psi _{1} &=&\left[
E+M-\Delta \left( r\right) \right] \psi _{2},  \label{dirac3b}
\end{eqnarray}%
where $\Sigma \left( r\right) =V\left( r\right) +S\left( r\right) $ and $%
\Delta \left( r\right) =V\left( r\right) -S\left( r\right) $.

If one adopts the following decomposition
\begin{equation}
\left(
\begin{array}{c}
\psi _{1} \\
\psi _{2}%
\end{array}%
\right) =\frac{1}{\sqrt{r}}\left(
\begin{array}{c}
\sum\limits_{m}f_{m}(r)\;\mathrm{e}^{im\varphi } \\
i\sum\limits_{m}g_{m}(r)\;\mathrm{e}^{i(m+s)\varphi }%
\end{array}%
\right) ,  \label{ansatz}
\end{equation}%
with $m+1/2=\pm 1/2,\pm 3/2,\ldots $, with $m\in \mathbb{Z}$, and inserting
this into Eqs. (\ref{dirac3a}) and (\ref{dirac3b}), we obtain%
\begin{eqnarray}
&&\left[ \frac{d}{dr}+\frac{s\left( m+s\right)-\frac{1}{2} }{r}-esA_{\varphi }\right]
g_{m} =\left[ E-M-\Sigma \right] f_{m} ,  \label{dirac4} \\
&&\left[ -\frac{d}{dr}+\frac{sm+\frac{1}{2}}{r}-esA_{\varphi }\right] f_{m} =%
\left[ E+M-\Delta \right] g_{m} .  \label{dirac5}
\end{eqnarray}
\noindent Note that it is impossible to obtain the equation (\ref{dirac5}) [or equation (\ref{dirac4})] from equation (\ref{dirac4}) [or equation (\ref{dirac5})] by mean of a charge conjugation or discrete chiral transformation, as already was uncovered in Ref. \cite{PRC73:054309:2006,IJMPE16:3002:2007,AP356:83:2015} for $1+1$ dimensions, and in Ref. \cite{PRC86:052201:2012} for $3+1$ dimensions.

For $\Sigma(r)=0$ [or $\Delta(r)=0$] with $E\neq M$ [or $E\neq-M$], the searching for solutions can be formulated as a Sturm--Liouville problem for the component $g(r)$ [or $f(r)$] of the Dirac spinor, as done in Ref. \cite{EPJA52:201:2016} for bound states. On the other hand, the solutions for $\Delta(r)=0$ with $E=-M$ and $\Sigma(r)=0$ with $E=M$, excluded from the Sturm--Liouville problem, can be obtained directly from the first--order equations (\ref{dirac4}) and (\ref{dirac5}), such solutions are called isolated solutions \cite%
{EPL108:30003:2014,AP338:278:2013,PS77:045007:2008,IJMPE16:2998:2007,IJMPE16:3002:2007, JPA40:263:2007,PS75:170:2007,PLA351:379:2006}.  

\section{Isolated solutions for the Dirac equation in $2+1$ dimensions}
\label{sec3}

For $\Sigma \left( r\right) =0$ with $E=M$, the first--order equations (\ref{dirac4}) and (\ref{dirac5}) reduce to
\begin{eqnarray}
&&\left[ \frac{d}{dr}+\frac{s(m+s)-\frac{1}{2}}{r}-seA_{\varphi }\right] g_{m}=0,
\label{diracme} \\
&&\left[ -\frac{d}{dr}+\frac{sm+\frac{1}{2}}{r}-seA_{\varphi }\right] f_{m}=2\left(
M-V \right) g_{m},  \label{diracma}
\end{eqnarray}%
whose general solution is
\begin{eqnarray}
g_{m}(r) &=&a_{+}r^{-s\left( m+s\right)+\frac{1}{2} }\mathrm{e}^{sev(r)},
\label{Mma} \\
f_{m}(r) &=&\left[ b_{+}-a_{+}I(r)\right] r^{sm+\frac{1}{2}}\mathrm{e}^{-sev(r)},
\label{Mmb}
\end{eqnarray}%
where $a_{+}$ and $b_{+}$ are normalization constants, and 
\begin{equation}\label{e1}
I(r)=\int dr\left[ 2M-2V\left( r\right) \right]r^{-(2sm+1)} \mathrm{e}^{2sev(r)}\,,
\end{equation}%
\begin{equation}\label{e2}
v(r)=\int^{r} A_{\varphi }(x)dx\,.
\end{equation}
\noindent Note that this sort of isolated solution cannot describe scattering states and is subject to the normalization condition
\begin{equation}
\int_{0}^{\infty }\left( |f_{m}(r)|^{2}+|g_{m}(r)|^{2}\right) dr=1.
\label{eq:norm}
\end{equation}
\noindent Because $f_{m}(r)$ and $g_{m}(r)$ are normalize functions, the possible isolated solution presupposes $A_{\varphi}\neq0$. This fact clearly shows that the normalization of the isolated solution is decided by the behavior of $v(r)$, i.e. the presence of $A_{\varphi}$ is an essential ingredient for the normalization of the isolated solution. Observing (\ref{Mma}) and (\ref{Mmb}), one can conclude that it is impossible to have both nonzero components simultaneously as physically acceptable solutions.

For $\Delta \left(r\right) =0$ with $E=-M$, the first--order equations (\ref{dirac4}) and (\ref{dirac5}) reduce to
\begin{align}
& \left[ \frac{d}{dr}+\frac{s\left( m+s\right)-\frac{1}{2} }{r}-esA_{\varphi } \right] g_{m}=-2\left[ M+V \right]f_{m}, \\
& \left[ -\frac{d}{dr}+\frac{sm+\frac{1}{2}}{r}-esA_{\varphi } \right]f_{m} =0,
\end{align}%
whose general solution is
\begin{eqnarray}
f_{m}(r) &=&a_{-}r^{sm+\frac{1}{2}}\mathrm{e}^{-sev(r)},  \label{fa} \\
g_{m}(r) &=&\left[ b_{-}-a_{-}H(r)\right] r^{-s\left( m+s\right)+\frac{1}{2} }\mathrm{e}^{sev(r)},  \label{ga}
\end{eqnarray}%
where $a_{-}$ and $b_{-}$ are normalization constants, and 
\begin{equation}
H(r)=\int dr\left[ 2M-2V\left( r\right) \right]r^{2sm+1} \mathrm{e}^{-2sev(r)}.
\end{equation}
\noindent The same conclusions obtained from the case $\Sigma \left( r\right) =0$ with $E=M$ are valid, i.e. the presence of $A_{\varphi }$ is an essential ingredient for the normalization of the isolated solution and it is impossible to have both nonzero components simultaneously.

Having set up the Dirac equation in $2+1$ dimensions, we are now in a position to use the machinery developed above in order to find isolated solutions with some specific forms for the external interactions. 
Assuming the external interactions as in Ref. \cite{EPJA52:201:2016},
\begin{equation}\label{vectorA}
\vec{A}=\left(0,\frac{B_{0}r}{2}+\frac{\Phi_{AB} }{2\pi r},0\right),  
\end{equation}%
\begin{equation}\label{vr}
V\left( r\right) =a\,r^{2}+br-\frac{c}{r},  
\end{equation}%
\noindent where $B_{0}$ is the magnetic field magnitude, $\Phi_{AB}$ is the flux parameter, $a$, $b$ and $c$ are constants.   

In this case, substituting  (\ref{vectorA}) in (\ref{e2}) one finds 
\begin{equation}\label{int}
v(r)=\frac{B_{0}r^{2}}{4}+\frac{\Phi_{AB}}{2\pi}\ln r\,. 
\end{equation}%
\noindent Now, using (\ref{int}) we are now in a position to find the isolated solutions for $\Sigma(r)=0$ with $E=M$ and $\Delta(r)=0$ with $E=-M$.

\subsection{Isolated solution for $\Sigma \left( r\right) =0$ with $E=M$}

In this case, the solutions (\ref{Mma}) and (\ref{Mmb}) become
\begin{eqnarray}
g_{m}(r) &=&a_{+}r^{s\left(\lambda-m\right)-\frac{1}{2}}\mathrm{e}^{s\delta r^{2}},
\label{Mma2} \\
f_{m}(r) &=&\left[ b_{+}-a_{+}I(r)\right] r^{-s\left(\lambda-m\right)+\frac{1}{2}}\mathrm{e}^{-s\delta r^{2}},
\label{Mmb2}
\end{eqnarray}%
\noindent where $\lambda=\frac{e\Phi_{AB}}{2\pi}$ and $\delta=\frac{eB_{0}}{4}$. In this case, for 
$\lambda>0$, $\delta>0$ and $s=1$, a normalizable solution is possible only for $a_{+}=0$. Therefore,
\begin{equation}\label{isol1}
\left(
\begin{array}{c}
f_{m} \\
g_{m}%
\end{array}%
\right) =b_{+}r^{m-\lambda+\frac{1}{2}}\mathrm{e}^{-\delta r^{2}}\left(
\begin{array}{c}
0 \\
1%
\end{array}%
\right)\,, 
\end{equation}%
\noindent independently of $a$, $b$, $c$ and $M$. Note that (\ref{isol1}) is square--integrable at the origin and satisfies $g_{m}(0)=0$ when $m-\lambda+\frac{1}{2}>0$. 

In the case, for $\lambda>0$, $\delta>0$ and $s=-1$, a normalizable solution requires $b_{+}=0$, and a good behavior of $I(r)$. For the Killingbeck potential (\ref{vr}), $I(r)$ can be expressed in terms of the incomplete gamma function \cite{ABRAMOWITZ1965} 
\begin{equation}
\gamma \left( \alpha,r\right) =\int\nolimits_{0}^{r}dt\,e^{-t}t^{a-1},\textrm{ }%
\textrm{Re }\alpha>0\,.
\end{equation}%
\noindent Because $\gamma\left( \alpha,r\right)$ tends to $\Gamma(\alpha)$ as $r\rightarrow\infty$, $f_{m}$ is not, in general, a square--integrable function.  So, a normalizable solution occurs when $M=a=b=c=0$ ($I(r)=0$). Therefore,
\begin{equation}\label{isol2}
\left(
\begin{array}{c}
f_{m} \\
g_{m}%
\end{array}%
\right) =a_{+}r^{m-\lambda-\frac{1}{2}}\mathrm{e}^{-\delta r^{2}}\left(
\begin{array}{c}
1 \\
0%
\end{array}%
\right)\,.
\end{equation}%
\noindent Note that in this case, the solution (\ref{isol2}) is square--integrable at the origin and satisfies $f_{m}(0)=0$ when $m-\lambda-\frac{1}{2}>0$. 

\subsection{Isolated solution for $\Delta \left(r\right) =0$ with $E=-M$}

In this case, the solutions (\ref{fa}) and (\ref{ga}) become
\begin{eqnarray}
f_{m}(r) &=&a_{-}r^{-s\left(\lambda-m\right)+\frac{1}{2}}\mathrm{e}^{-s\delta r^{2}},  \label{fa2} \\
g_{m}(r) &=&\left[ b_{-}-a_{-}H(r)\right] r^{s\left(\lambda-m\right)-\frac{1}{2} }\mathrm{e}^{s\delta r^{2}}\,.  \label{ga2}
\end{eqnarray}%
\noindent Following the same procedure of the previous case, for $s=-1$ a normalizable solution occurs when $a_{-}=0$. The isolated solution is given by
\begin{equation}\label{isol3}
\left(
\begin{array}{c}
f_{m} \\
g_{m}%
\end{array}%
\right) =b_{-}r^{m-\lambda-\frac{1}{2}}\mathrm{e}^{-\delta r^{2}}\left(
\begin{array}{c}
0 \\
1%
\end{array}%
\right)\,, 
\end{equation}%
\noindent with $m-\lambda-\frac{1}{2}>0$. 

For $s=1$ a normalizable solution is possible only for $b_{-}=M=a=b=c=0$. In this case, the normalizable solution is given by
\begin{equation}\label{isol4}
\left(
\begin{array}{c}
f_{m} \\
g_{m}%
\end{array}%
\right) =a_{-}r^{m-\lambda+\frac{1}{2}}\mathrm{e}^{-\delta r^{2}}\left(
\begin{array}{c}
1 \\
0%
\end{array}%
\right)\,, 
\end{equation}%
\noindent with $m-\lambda+\frac{1}{2}>0$.

\section{Quadratic form of the Dirac equation in $2+1$ dimensions}
\label{sec4}

Now, we investigate the dynamics for $E\neq \pm M$. For this, we
choose to work with Eq. (\ref{dirac2}) in its quadratic form. After
application of the operator%
\begin{equation}
\beta \left[ \left( M+S\left( r\right) \right) +\beta \left( E-V\left(
r\right) \right) +\boldsymbol{\gamma }\cdot \boldsymbol{\pi }\right] ,
\end{equation}%
we get \cite{EPJC75:321:2015}%
\begin{eqnarray}
&&\left\{ \vec{p}^{2}-2e\left( \vec{A} \cdot \vec{p}\right) +e^{2}\left( \vec{A}\right) ^{2}\right\} \psi \left( \mathbf{r}\right)  \notag \\
&&+\left\{ \left[ M+S\left( r\right) \right] ^{2}-\left[ E-V\left( r\right) %
\right] ^{2}-e\vec{\sigma }\cdot \vec{B} \right\} \psi \left( \mathbf{r}\right)  \notag \\
&&-\left( \frac{\partial S\left( r\right) }{\partial r}\sigma _{2}+i\frac{%
\partial V\left( r\right) }{\partial r}\sigma _{1}\right) \psi \left(
\mathbf{r}\right) =0.  \label{diracB}
\end{eqnarray}
In this stage, it is worthwhile to mention that the Eq.~(\ref{diracB}) is
the correct quadratic form of the Dirac equation with minimal, vector and
scalar couplings, because the Pauli term is considered.

Now, we focus attention on a misconception diffused in the literature. The vector potential in (\ref{vectorA}) furnishes one magnetic field perpendicular to the plane $\left( r,\varphi \right) $, given by \cite{PRL64:503:1990,PRD48:5935:1993,JMP42:1933:2001}%
\begin{equation}\label{cmcd}
\vec{B}=\vec{\nabla}\times\vec{A}=\left(B_{0}+\frac{\Phi_{AB}\delta(r)}{2\pi r}\right)\vec{\hat{z}}\,,
\end{equation}%
\noindent and not simply $\vec{B}=B_{0}\vec{\hat{z}}$ as considered in the Refs. \cite{ChPB21:110302:2012,PB407:4523:2012,FBS54:1987:2013,MPLB27:1350176:2013,
ADHEP2013:562959:2013,ADHEP2013:491648:2013,AP341:153:2014,AP353:282:2015,EPJA52:201:2016,PRE93:053201:2016}.
Here, we can interpret the first term in (\ref{cmcd}) as an constant external magnetic field and the second term as the magnetic field produce by a solenoid. If the solenoid is extremely long, the
field inside is uniform, and the field outside is zero. However, in a
general dynamics, the particle is allowed to access the $r=0$ region. In
this region, the magnetic field is non-null. If the radius of the solenoid
is $r_{0}\approx 0$, then the relevant magnetic field is $\vec{B}\sim
\delta (r)$. Therefore, on the study of the dynamics of a particle with spin, such term cannot be neglected in the equation of motion \cite{PRL64:503:1990}, because has important implications on the physical quantities of interest, such as energy eigenvalues, scattering matrix and phase shift (see Ref. \cite{AP339:510:2013} for more details). This situation has not been
accomplished in Refs. \cite{AP341:153:2014,EPJA52:201:2016}. 

\subsection{Exact spin symmetry limit: $S(r)=V(r)$}
\label{sec4:1}

By using the condition $S(r)=V(r)$ ($\Delta(r)=0$) in (\ref{diracB}), we obtain a second order differential equation for $\psi_{1}$. In this case, the upper component of the Dirac spinor can be considered as
\begin{equation}\label{solsm}
\psi _{1}=\sum\limits_{m}\frac{f_{m}(r)}{\sqrt{r}}\;\mathrm{e}^{im\varphi }.  
\end{equation}%
\noindent So, substituting (\ref{vectorA}), (\ref{vr}), (\ref{cmcd}) and (\ref{solsm}) in (\ref{diracB}), the equation for $f_{m}\left(r\right) $ becomes%
\begin{equation}\label{eigen}
Hf_{m}\left( r\right) =k^{2}f_{m}\left( r\right) ,  
\end{equation}%
\noindent with%
\begin{equation}\label{hfull}
H=H_{0}-\frac{es\Phi_{AB}\delta (r)}{2\pi r},  
\end{equation}%
\begin{equation}\label{hzero}
H_{0}=-\frac{d^{2}}{dr^{2}}+\eta\,r^{2}+\rho\,r+\frac{\nu }{r^{2}}-\frac{\mu}{r}\,,  
\end{equation}%
\noindent where%
\begin{eqnarray}
\eta &=&2\left( E+M\right)a+\frac{e^{2}B_{0}^{2}}{4} , \\
\rho &=& 2\left( E+M\right)b , \\
\nu &=&\left( m-\frac{e\Phi_{AB}}{2\pi} \right) ^{2}-\frac{1}{4} , \\
\mu &=& 2\left( E+M\right)c , \\
k^{2} &=& E^{2}-M^{2}+e\left( m+s \right)B_{0}-\frac{e^{2}B_{0}\Phi_{AB}}{2\pi}  .
\end{eqnarray}
\noindent Note that the equation (\ref{eigen}) depends on the spin projection $s$ and it is different from that given in Ref. \cite{EPJA52:201:2016}. It is worthwhile to mention that (\ref{eigen}) is the correct equation of motion for the upper component of the Dirac spinor under the exact spin symmetry limit. 

The solution considering only the term $H_{0}$ (\ref{hzero}), with $\eta$ necessarily real and positive, is the solution of the Schr\"{o}\-dinger equation for the three--dimensional harmonic oscillator plus a Cornell potential \cite{PRC86:052201:2012}, which can be obtained by recurring to the properties of the biconfluent Heun equation. This potential was considered in Refs. \cite{JPA19:3527:1986,RONVEAUX1995}, but the authors misunderstood the full meaning of the potential and made a few erroneous calculations.

The presence of a $\delta(r)$ interaction in the radial Hamiltonian (\ref{eigen}) makes the problem more complicated to resolve. For this case, the most adequate procedure to address this problem is by means of the self--adjoint extension approach \cite{JMP26:2520:1985}, but unhappily the self--adjoint extension for a biconfluent Heun equation is unknown.

\subsection{Exact pseudo--spin symmetry limit: $S(r)=-V(r)$}
\label{sec4:2}

By using the condition $S(r)=-V(r)$ ($\Sigma(r)=0$) in (\ref{diracB}), we obtain a second order differential equation for $\psi_{2}$. In this case, the lower component of the Dirac spinor can be considered as
\begin{equation}
\psi _{2}=i\sum\limits_{m}\frac{g_{m}(r)}{\sqrt{r}}\;\mathrm{e}^{i(m+s)\varphi }.  \label{solpsm}
\end{equation}
\noindent So, substituting (\ref{vectorA}), (\ref{vr}), (\ref{cmcd}) and (\ref{solpsm}) in (\ref{diracB}), the equation for $g_{m}\left(r\right) $ becomes%
\begin{equation}\label{eigen2}
\tilde{H}g_{m}\left( r\right) =\tilde{k}^{2}g_{m}\left( r\right) ,
\end{equation}%
with%
\begin{equation}
\tilde{H}=\tilde{H}_{0}-\frac{es\Phi_{AB}\delta (r)}{2\pi r},
\end{equation}%
\begin{equation}
\tilde{H}_{0}=-\frac{d^{2}}{dr^{2}}+\tilde{\eta}\,r^{2}+\tilde{\rho}\, r+\frac{\tilde{\nu} }{r^{2}}-\frac{\tilde{\mu}}{r}\,,  
\end{equation}%
\noindent where%
\begin{eqnarray}
\tilde{\eta} &=&2\left( E-M\right)a+\frac{e^{2}B_{0}^{2}}{4} , \\
\tilde{\rho} &=& 2\left( E-M\right)b , \\
\tilde{\nu} &=&\left( m+s-\frac{e\Phi_{AB}}{2\pi} \right) ^{2}-\frac{1}{4} , \\
\tilde{\mu} &=& 2\left( E-M\right)c , \\
\tilde{k}^{2} &=& E^{2}-M^{2}+e\left(m+2s\right) B_{0}-\frac{e^{2}B_{0}\Phi_{AB}}{2\pi}  .
\end{eqnarray}
\noindent The equation (\ref{eigen2}) is the correct equation of motion for the lower component of the Dirac spinor under the exact pseudo--spin symmetry limit and again it is different from that given in Ref. \cite{EPJA52:201:2016}. Analogously to the previous case, (\ref{eigen2}) depends on the spin projection $s$.

\section{Final remarks}
\label{sec5}

In this paper, we reinvestigated the issue of the Dirac equation in $2+1$ dimensions with a Killingbeck radial potential including an external magnetic and Aharonov--Bohm (AB) flux fields. Using a adequate representation for the Dirac matrices, we solved the first order Dirac equation and found solutions for $\Sigma(r)=0$ with $E=M$ and $\Delta(r)=0$ with $E=-M$, which are called isolated solutions because they are excluded from the Sturm--Liouville scheme. We showed that these solutions depend on the spin projection parameter $s$ and that the presence of the vector potential is indispensable for a normalizable isolated solution. Also, we pointed out a misleading treatment recently propagated in the literature with respect to Aharonov--Bohm (AB) potential. Finally, we also showed the correct quadratic form of the Dirac  equation in $2+1$ dimensions taking into account the spin and pseudo--spin symmetries, which includes a $\delta(r)$ function as a consequence of the Pauli term. It is known that to properly study the dynamics of the particle in this case, the most adequate procedure is the self--adjoint extension approach \cite{JMP26:2520:1985}, but unhappily the self--adjoint extension for a biconfluent Heun equation is unknown. This last problem is open.  

\begin{acknowledgement}
This work was supported in part by means of funds provided by CNPq (grants
455719/2014-4 and 304105/2014-7). Angel E. Obispo thanks to CAPES for support through a scholarship under the CAPES/PNPD program.
\end{acknowledgement}


\end{document}